\def\year{2015}
\documentclass[letterpaper]{article}
\usepackage{aaai}
\usepackage{times}
\usepackage{helvet}
\usepackage{courier}
\usepackage{graphicx}
\begin{document}
\title{Local Variation of Collective Attention in Hashtag Spike Trains}
\author{Ceyda Sanli \and Renaud Lambiotte\\
CompleXity and Networks, naXys, Department of Mathematics\\
University of Namur, 5000 Namur, Belgium\\
}

\maketitle
\begin{abstract}
\begin{quote}
In this paper, we propose a methodology quantifying temporal patterns of nonlinear hashtag time series. Our approach is based on an analogy between neuron spikes and hashtag diffusion. We adopt the local variation, originally developed to analyze local time delays in neuron spike trains. We show that the local variation successfully characterizes nonlinear features of hashtag spike trains such as burstiness and regularity. We apply this understanding in an extreme social event and are able to observe temporal evaluation of online collective attention of Twitter users to that event.
\end{quote}
\end{abstract}

\section{Introduction}
\noindent Hashtag diffusion in Twitter social network is nonlinear in time. Pairwise or higher order temporal correlations, bursts, and regular patterns are observed in data analysis. The distribution of time delays between two successive hashtag activities gives a power-law scaling with fat tails~\cite{scientificrumor}, on the contrary to an exponential distribution suggested for an independent Poisson process. A potential reason addressed is that earlier hashtags influence coming hashtags such that past hashtags can both cooperate and compete with present hashtags~\cite{cooperationandcompetition,Competition_memes}. Heterogeneity of individual online user behavior in micro scale and self-organized cascades~\cite{leskovec} due to unequal selection~\cite{onlinepopularityheterogeneity,Competition_memes_limitedattention,Competition_inducedcriticality,Competition_memes,Competition_advertisement,2015arXiv150105956G} in the hashtag pool in macro scale, and the underlying cyclic rhythm of twitting habit~\cite{burstydynamicsTwitter,2014arXiv1411.0722F,2015arXiv150203224M,2015arXiv150303349S} are further factors driving time-dependent hashtag propagation. Although preserving highly nonlinear nature, building tools to characterize hashtag time-series, except obtaining the distribution functions, has not been considered in detail, yet.

\begin{figure}[Ht!]
\begin{center}
\includegraphics[width= 7.5cm]{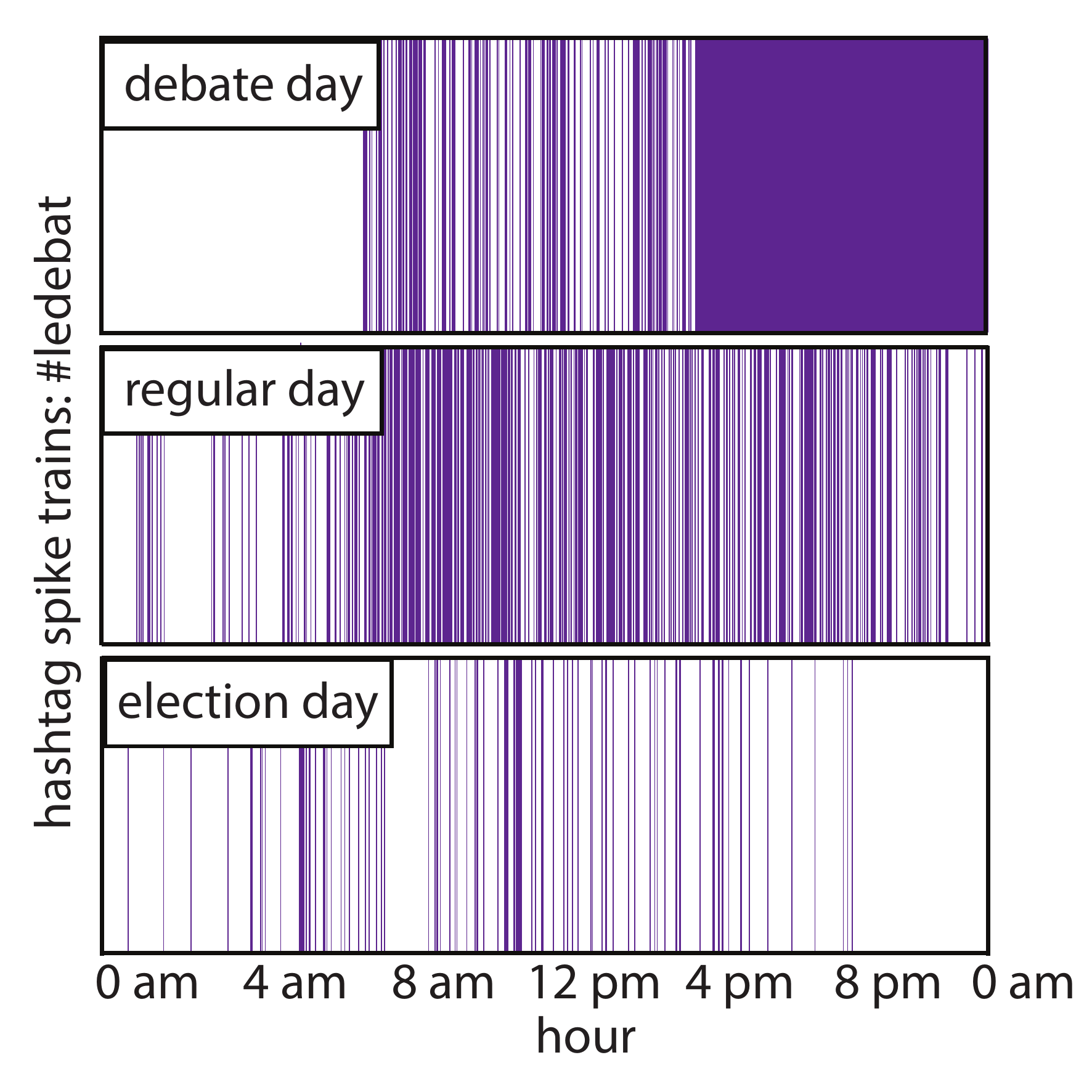}%{Fig1_dailytweetComp.pdf}
\caption{Hashtag spike trains of $\#$ledebat in different days covering extreme social events such as the debate of the French presidential election-2012 held on May 2 and the election held on May 6, and a regular day between them, e.g May 4. The upper row represents the dynamics in the debate day. Collective attention during the debate gives tremendous amount of activity on the hashtag and so we observe a continuous series on the contrary to the distinguished spikes before 4 pm. The middle row is for a regular day after the debate, followed by the spike train in the election day in the below row. A decay in the activity on the hashtag $\#$ledebat is visible from the top to the bottom and the process suggests highly nonlinear characteristics in each day.
\label{fig:combineLv}}
\end{center}
\end{figure}

Extreme social events such as elections and protests~\cite{spanishprotestPLoSONE,spanishprotestSciRep}, announcement of scientific innovations~\cite{scientificrumor}, and panic events such as crisis~\cite{determinecrisisviahash} and earthquakes~\cite{earthquakeTwittercollectiveattention} artificially deform Twitter network and encourage massive amount of hashtag activity in a short time window, as shown in Figure 1. The resultant emergent online behavior is both empirically~\cite{yangleskovec_temporalpatterns,2015arXiv150203224M} and theoretically~\cite{2015arXiv150203224M} studied and distinct temporal properties of collective attention are quantified. These properties are significantly important to be able to predict these extreme, but rare social events~\cite{determinecrisisviahash,Altmann_extreme_events}.  

Our main motivation is to establish a systematic methodology to distinguish real noisy hashtag signals to independent random signals and to extract temporal patterns from the real signals. We apply an approach called the local variation $L_V$, originally introduced to analyze noisy neuron spike trains and to detrend for salient dynamics of neurons~\cite{Lv1,Lv3,Lv5}. After convincing the usage of $L_V$ in semantic analysis, which has been performed extensively in our recent work~\cite{2015arXiv150303349S}, we present a promising study on evaluation of collective attention by performing $L_V$ on a political election. Remarkable difference in $L_V$ in rush period suggests that local nonlinear features could predict extreme social events.

\section{Data Set}
\subsection{Data Collection}
\noindent The data is collected via publicly open Twitter API. A fine time window, between April 30, 2012 and May 10, 2012, is determined on purpose to be able to cover two social events such as the political debate on the French presidential election-2012 held on May 2 and the election day held on May 6. Having 10 days data helps us to visualize activity in regular days, both between and after these extreme events, and compare the difference in hashtag dynamics. During this period, all twitting activity, but only the users addressed in France is considered not to deal with time differences between countries and regions and other potential social events held on in the same period. The time resolution is 1 second and no language selection is applied. 

We examine 295,697 unique hashtags out of 2,942,239 tweets include at least one hashtag, which is $30\%$ of all tweets. 228,525 online users, almost half of the total online users, are associated with hashtag diffusion. The network in the period contains hashtags directly related to the debate, election, and two candidates Francois Hollande and Nicolas Sarkozy for the presidency of France. Ranking them by the number of appearance (frequency) or equivalently popularity $p$, from the most popular to the least, we have $\#$ledebat (180946), $\#$hollande (143636), $\#$sarkozy (116906), $\#$votehollande (99908), $\#$avecsarkozy (67549), $\#$ledebat (66668) [in French], $\#$france2012 (20635), $\#$presidentielle (13799), and many others with lesser $p$. The numbers inside the parenthesis present the corresponding $p$. These popular hashtags are at the top of the others in the pool, e.g. $0.0001\%$ of all hashtags.

\subsection{Real Hashtag Spike Trains}
\noindent Single hashtag diffusion in time can be represented as a spike train, as shown in Figure 1. Each spike represents that the corresponding hashtag used at that time without specifying ways and users. Having the resolution 1 second, the spike time of multiple events occurring in a second cannot be distinguished and therefore in this situation only one appearance is counted. We construct spike trains for all hashtags observed in the data ordering from the earliest appearance time to the latest time, e.g. $\ldots,\,\tau_{i-1},\,\tau_i,\,\tau_{i+1},\,\ldots\:$. Each hashtag has a unique number of (exact) appearance, popularity $p$. 

\subsection{Randomized Hashtag Spike Trains}
\noindent To be able to compare real dynamics with an artificial and independent one, the randomized version of real hashtag spike trains is established serving as a null model. First, all spikes coming from any hashtags are combined, giving a single merged hashtag spike train. Uniforming spike appearance, one spike at a spike time, is still valid. Children randomized hashtag spike trains are obtained by uniformly permuting the matrix $T$ of the spike times of the merged train by $p$ times, the number of spikes of the desired real train we compare. We apply randperm($T$, $p$) in Matlab and have $p$ times uniformly distributed unique independent random spike times, e.g. $\ldots,\,\tau^r_{i-1},\,\tau^r_i,\,\tau^r_{i+1},\,\ldots\:$.

\section{Local Variation}
The local variation $L_V$, specifically defined to quantify nonlinear neural time-series and to uncover temporal patterns in neuron spike trains, is defined at spike time $\tau_i$~\cite{Lv5}
\begin{equation}
L_V=\frac{3}{N-2}\sum\limits_{i=2}^{N-1} \left(\frac{\Delta\tau_{i+1}-\Delta\tau_{i}}{\Delta\tau_{i+1}+\Delta\tau_{i}}\right)^2,
\label{Eq:Lv_Deltatau}
\end{equation}
$\Delta\tau_{i+1}$ = $\tau_{i+1}-\tau_i$ and $\Delta\tau_{i}$ = $\tau_{i}-\tau_{i-1}$.  $\Delta\tau_{i+1}$ quantifies forward delay and $\Delta\tau_{i}$ represents backward waiting time. Importantly, the denominator normalizes the quantity such as to account for local variations of the rate at which events take place. 

By definition, $L_V$ takes values in the interval [0:3]. Furthermore, it is derived that $L_V$ is on average equal to 1, $\langle L_V\rangle$ = 1, if the underlying process described by an independent Poisson distribution, which the distribution of the inter-spike intervals gives an exponential function~\cite{Lv1}. Here, the brackets describe the average taken over the given distribution. All other situations can be generalized by Gamma processes~\cite{Lv1,Lv3} and $\langle L_V\rangle$ should be significantly different than 1. For instance, $\langle L_V\rangle\approx3$ if the hashtag spike trains are extremely bursty (irregular), on the other hand $\langle L_V\rangle\approx0$ while the trains present regular (homogeneous) temporal patterns~\cite{2015arXiv150303349S}.

\begin{figure}[Ht!]
\begin{center}
\includegraphics[width= 8.0cm]{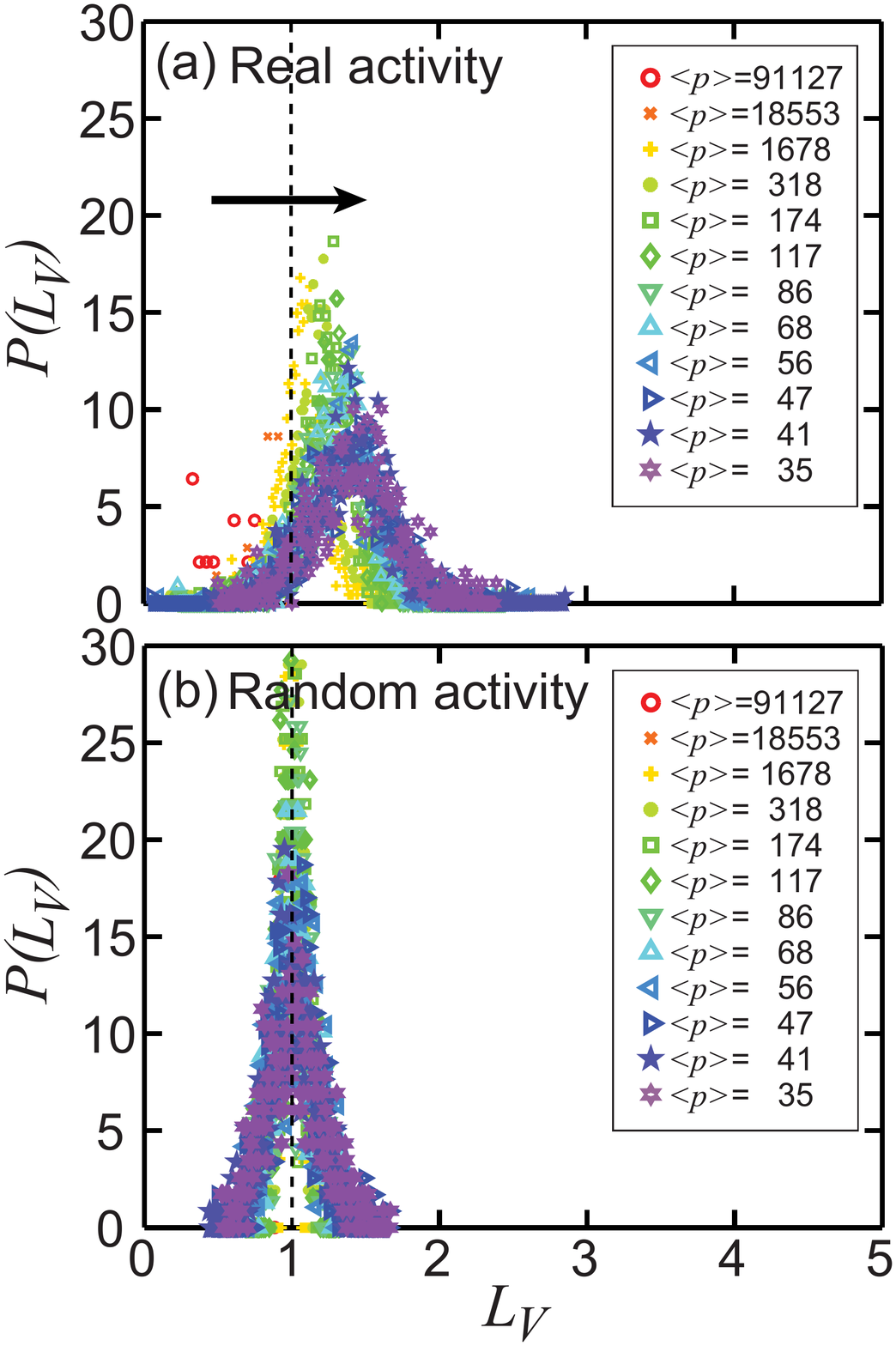}%{Fig1_dailytweetComp.pdf}
\caption{Probability density function of local variation $L_V$, $P(L_V)$, of hashtag spike trains~\cite{2015arXiv150303349S}. (a) Real hashtag spike trains. We observe a clear shift, to the higher values of $L_V$, in the peak positions while decreasing hashtag popularity $p$, which indicates that the process becomes bursty (irregular). In any $p$, the mean values never gives 1, none of the real signal is Poisson process. (b) Randomized hashtag spike trains. Independent of $p$, all curves suggest fluctuations around 1, as expected for temporarily independent signals. To satisfy a better visualization, the results are grouped based on ranking $p$ from the most popular to the least popular ones: High $p$, red and orange symbols, moderate $p$, yellow and green symbols, and low $p$, blue and purple symbols. 
\label{fig:combineLv}}
\end{center}
\end{figure}

Figure 2 shows the results of our $L_V$ analysis, for both real and randomized hashtag spike trains. The probability distribution of $P(L_V)$ of the calculated values of $L_V$ on the two data sets, with classifying hashtag groups in popularity $p$, presents distinct behavior. Whereas $\langle L_V\rangle$ = 1 for any groups of $p$ for the randomized trains, suggesting Poisson processes, $\langle L_V\rangle$ never indicates 1 for the real trains. The randomization dampens nonlinearity of the real trains, temporal correlations, burstiness, and regularity in series and construct \textit{statistically} stationary and independent processes, yet time-dependent events. Therefore, we characterize time-dependent Poissons in Figure 2(b), $P(L_V)$ fluctuates around 1. However, all nonlinearities are present in the real data, and so in $P(L_V)$. Describing regular patterns for popular hashtags (high $p$), red and orange symbols, the trains become bursty (irregular) due to local temporal correlations for moderate, yellow and green symbols, and for low popularity, blue and purple symbols. The trend is captured in the shift of the peak positions of $P(L_V)$ from small $L_V$ to large $L_V$ decreasing $p$ in Figure 2(a). Consequently, we find that not only for neurons but also for hashtags $L_V$ is a successful tool to characterize salient dynamics in nonlinear social time-series.

\section{Empirical Application: Collective Attention}
\noindent We now utilize $L_V$ for more practical purposes and ask: Can $L_V$ predict extreme social events? Our investigation will be presented below is far from a complete understanding. However, we will be able to capture temporal evaluation of online emergent behavior as a result of collective attention of twitting on the French presidential election-2012, in the first week of May 2012. 

We specifically compare hashtag diffusion in extreme days, the debate day (May 2) and the election day (May 6) with the dynamics in a regular day between these events, e.g. May 4. Instead of considering all hashtags in the pool, as done in the previous Section, we concentrate on topic related hashtags such as $\#$ledebat (180946), $\#$hollande (143636), $\#$sarkozy (116906), $\#$votehollande (99908), $\#$avecsarkozy (67549), and $\#$ledebat (66668) [in French]. The numbers in the parenthesis indicate $p$ of the corresponding hashtag. 

Local variation $L_V$ is obtained for these topic-oriented hashtag spike trains. The trains are constructed separately for the three days. $L_V$ for each train and for each day is calculated considering time window with duration 1 hour.  Figure 3 presents the results in the debate (left), regular (middle), and election (right) days. The top row [Figure 3(a)] shows $L_V(t)$ in the days in hour resolution. The below row [Figure 3(b)] summarizes the twitting activity as the tweets including listed hashtags in the legend versus time, again in hour resolution. 

Rush hours in online communications during the debate and the announcement of the election result are highlighted in the shaded yellow rectangle and with the yellow vertical line, respectively. Significant decays in $L_V(t)$ for both the debate and election days, synchronizing perfectly with the peak of the counts, indicate regular activation of the online users on the discussion of the election and so describe no burstiness, $L_V(t)\approx0$. This trend is not observed at all for the regular day and mainly the cyclic rhythm of Twitter network~\cite{2015arXiv150303349S} characterize the values of $L_V(t)$. While large amount of fluctuations present in inactive hours [0 am:6 am], the rest of the day $\langle L_V(t)\rangle\approx1$ suggesting time-dependent Poisson processes. These results are preliminary, but promising since the stages of collective attention are clearly visible on $L_V(t)$.

\begin{figure*}[Ht!]
\begin{center}
\includegraphics[width= 18.0cm]{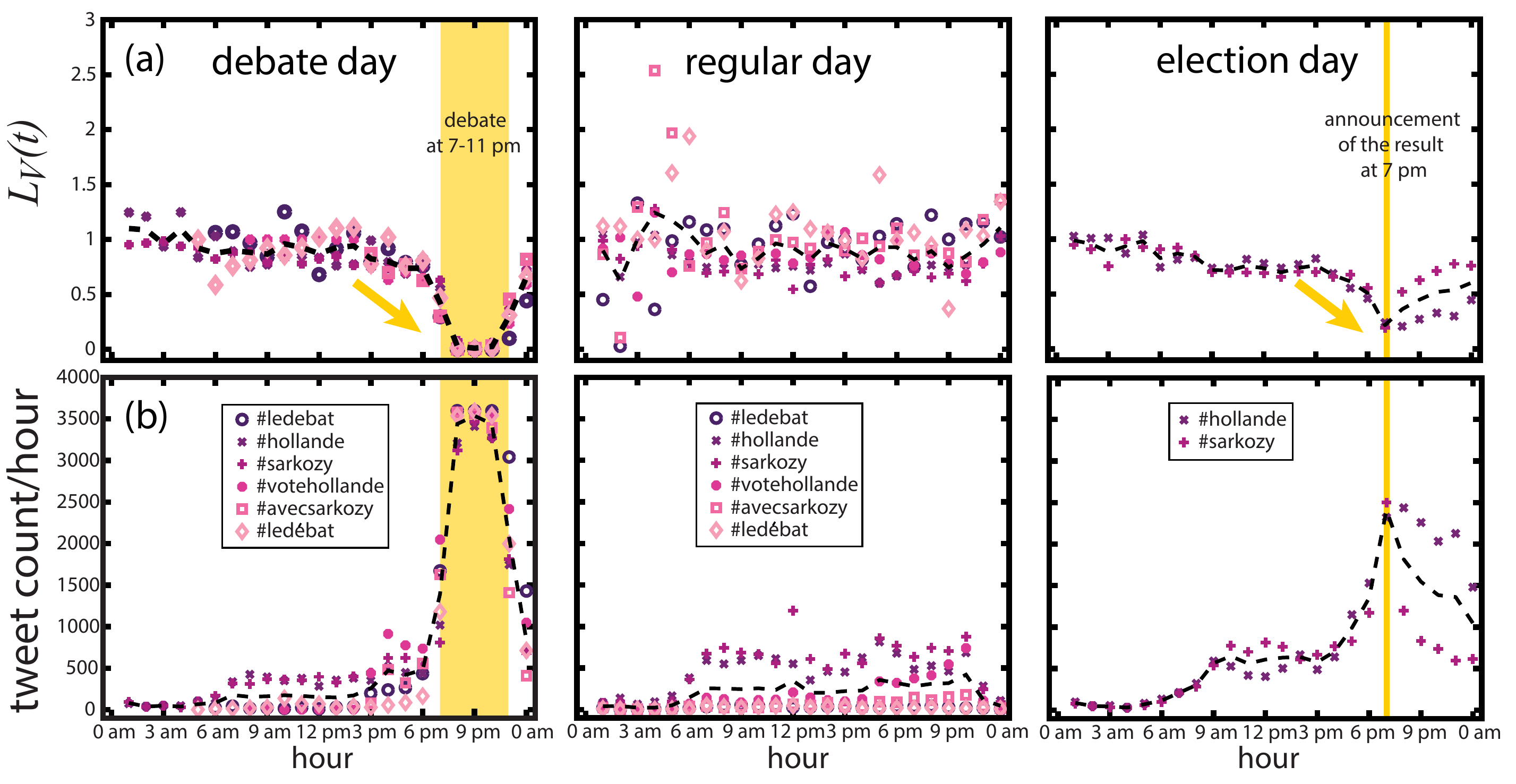}%{Fig1_dailytweetComp.pdf}
\caption{Characterizing temporal evaluation of collective attention. From left to right, the debate (May 2, 2012), a regular day (May 4, 2012), and the election day (May 6, 2012) are shown. (a) The local variation of $L_V(t)$ on the topic-related hashtags about the debate and the election. The shaded yellow rectangle covers the debate hours and the yellow line indicates the announcement of the election. Significant decays in $L_V(t)$, in left and right windows, match well with the schedule of the events. However, no remarkable trend is observed in the regular day (middle panel). (b) Counting tweets, including at least one of the hashtags addressed in the legends, per hour. The activity increase in time coincides successfully with the decays in $L_V(t)$, indicating that the collective attention homogenizes the hashtag propagation and so the hashtag spike trains in this limit present temporal regularity.
\label{fig:combineLv}}
\end{center}
\end{figure*}

\section{Discussion and Future Work}
\noindent The main purpose of this paper is to establish a tool for noisy social time-series and uncover nonstationary features and temporal patterns, specifically in an online emergent limit. Our comparative test on the real and randomized data sets shows that the local variation $L_V$, a metric introduced to quantify the fluctuations of neuron spike trains as compared to a local characteristic time, works successfully in hashtag spike trains, as well. This encourages us to develop further tools, for instance to predict extreme online events by evaluating the early noisy signal prior to an extreme event. As an example, we consider the week of the French presidential election-2012. This fine time window is well suitable for our aim and we find that $L_V$ is sensitive enough to distinguish collective attention period, users are active homogeneously in time, from the preceding period where temporal heterogeneity is present and therefore a prediction would be satisfied by performing better statistics in the decay of $L_V(t)$.

We obtain $L_V(t)$ is almost 0 in rush periods. Such artificial regularity originates from our assumption due to lack of time resolution below 1 second. Although we observe heterogeneity in hashtag spike trains in rush hours in the empirical data, uniforming spike appearance setting to 1 in any spike time creates unnatural homogeneity in emergent limit. To resolve this, the trains should be constructed preserving the heterogeneity in the data and so $L_V$ must be re-introduced for nonuniform number of spikes at different spike times in a train. 

\section{ Acknowledgments}
C. Sanli acknowledges supports from the EU 7th Framework OptimizR Project and FNRS. This paper presents research results of the Belgian Network DYSCO, funded by the Interuniversity Attraction Poles Programme, initiated by the Belgian State, Science Policy Office.
%\bigskip
%\noindent Thank you for reading these instructions carefully. We look forward to receiving your electronic files!

\bibliographystyle{aaai}
\bibliography{W3_refs_LvCollectiveAttention_SanliLambiotte}

\end{document}